# Infrared Diffuse Interstellar Bands in the Galactic Centre Region

T. R. Geballe, F. Najarro, D. F. Figer, B. W. Schlegelmilch, D. de la Fuente

**The spectrum of any star viewed through a sufficient quantity of diffuse interstellar material reveals a number of absorption features collectively called "diffuse interstellar bands" (DIBs). The first DIBs were reported 90 years ago[1], and currently well over 500 are known[2]. None of them has been convincingly identified with any specific element or molecule, although recent studies suggest that the DIB carriers are polyatomic molecules containing carbon[3,4,5]. Most of the DIBs currently known are at visible and very near-infrared wavelengths, with only two previously known at wavelengths beyond one micron (10,000 Å), the longer of which is at 1.318 μm[6]. Here we report the discovery of thirteen diffuse interstellar bands in the 1.5-1.8 μm interval on high extinction sightlines toward stars in the Galactic centre. We argue that they originate almost entirely in the Galactic Centre region, a considerably warmer and harsher environment than where DIBs have been observed previously. The relative strengths of these DIBs toward the Galactic Centre and the Cygnus OB2 diffuse cloud are consistent with their strengths scaling mainly with extinction by diffuse material.**

Figure 1 shows spectra of three Galactic Centre (GC) stars, GCS3-2[7], X174516[8], and qF362[9] in a small portion of the H-band (1.5-1.8 μm). These three stars have dramatically different intrinsic spectra, yet they have four common absorption features in this wavelength interval. These features are present in the spectra of all six GC stars that we have observed to date, whose spectral classifications range from late B to Wolf-Rayet. A weaker fifth feature is detected in most of the stars. Their full H-band spectra show considerable diversity, as expected, with the exception of these absorptions and a number of additional ones reported below. This is evidence that the newly discovered absorption features are formed in interstellar material. Further evidence, comes from their fixed wavelengths from star to star, in comparison to the wavelength shifts in the Brackett lines due to the radial velocities of the stars and in some cases their stellar winds. The strengths of the stronger features from star to star are the same to within 20 percent. GCS3-2 and qF362 are separated from each other on the plane of the sky by about 1 pc (we assume a GC distance of 8 kpc), whereas X174516 is separated from each of them by about 55 pc and thus is observed on a distinctly different sightline.

Searches of atomic line databases show that none of the absorption features corresponds within 0.002 μm to atomic or ionic lines with low-lying lower energy levels. The irregular pattern of their wavelengths and their differing intensities do not lend themselves to identification as one or more simple molecules. None has been reported in the spectra of young stars embedded in dense molecular clouds. However, previously unpublished spectra of bright stars in the Cygnus OB2 cluster contain four of the five absorption features (Fig. 1). The Cygnus OB2 cluster lies behind a diffuse cloud that attenuates optical light by 5-6 magnitudes[10] and is responsible for the numerous DIBs in the optical spectra of these stars[11], as well as the two previously discovered J-band DIBs[6,12]. Finally, as discussed below, in the J-band (1.1-1.35 μm) spectrum of qF362 (the only one of the three that we have observed in that band), the longer wavelength of the two previously discovered J-band DIBs[6] is clearly detected at 1.318 µm; the signal-to-noise ratio at the wavelength of the shorter wavelength J-band DIB is too low to yield a high-confidence detection.

Based on all of the above evidence, we conclude that the newly discovered features are members of the family of diffuse interstellar bands. They are the longest wavelength DIBs reported to date. Our longer wavelength K-band (2.0-2.4 µm) spectra of these stars do not reveal additional DIBs, to optical depths limits of ~0.02 or better over most of that band. Figure 2 shows spectra of the complete set of newly discovered H-band DIBs, and Table 1 lists their properties.

Where along the 8 kpc long sightline to the GC are the carriers of the H-band DIBs located? The dust along that sightline produces approximately 30 visual magnitudes of extinction[9], rendering the GC unobservable at optical wavelengths. Approximately one-third of the extinction occurs in dense molecular clouds, presumably located in intervening spiral arms, whereas the remaining two-thirds arises in diffuse clouds [13]. The newly discovered DIBs are thus associated with about 20 visual magnitudes of extinction. It is not surprising then, that they are considerably stronger toward the GC than toward Cygnus OB2.

There is strong evidence that the diffuse interstellar gas on the sightline to the GC is found, almost in its entirety, very close to the centre, in a region of radius ~200 pc known as the Central Molecular Zone (CMZ). Spectroscopy of the molecular ion $H_3^+$ toward GC sources, including GCS3-2, has shown that the CMZ contains a vast quantity of diffuse gas [14, 15]. Because the temperature of the gas in the CMZ, which can be measured using $H_3^+$, is much higher than in diffuse clouds in the Galactic plane, and because of the high velocity dispersion of the gas in the CMZ, (~150 km s$^{-1}$ in the direction of GCS3-2[14]) one can use high resolution spectroscopy of $H_3^+$ to clearly distinguish between diffuse material in and outside the CMZ. There is no evidence on these sightlines for a significant fraction of the diffuse cloud material being located outside of the CMZ. If, as we suspect, the H-band DIBs largely originate in the CMZ's diffuse gas, their widths are probably Doppler-broadened similarly to $H_3^+$. That broadening, corresponding ~0.0008 µm in the H-band, is greater than the resolution of the spectrum (0.0004 µm). Thus some of the newly discovered DIBs may have intrinsic spectral widths that are much less than they appear toward the GC, perhaps only a few Angstroms. This is not unusual, as many DIBs at visual wavelengths have widths of ~ 1 Å[16]. However, as the precise velocity distribution of the material producing these DIBs is not known, the laboratory wavelengths of the new bands currently should be considered accurate to only ±0.0010 µm (10 Å).

Spectroscopy of GC sources is much more difficult in the J-band than in the H-band, because the attenuation by dust is nearly an order of magnitude greater in the J-band. Nevertheless, we have secured a J-band spectrum of qF362, in which the previously discovered 1.318 µm DIB is prominently present. Figure 3 shows the portion of the spectrum containing this DIB. The strengths of the 1.318 µm and 1.527 µm DIBs in the GC sources are both about four times greater than they are in the spectra of stars in Cygnus OB2. If the strengths of these DIBs scale roughly with extinction[16], they imply that there are 20 mag of visual extinction in diffuse clouds toward the GC, which is consistent with previous estimates[13].

The diffuse interstellar medium in the GC is a considerably harsher environment than the diffuse clouds where DIBs have been previously observed. The gas temperatures in the centre are 200-300 K[13], compared to 30-100 K in Galactic diffuse clouds[18] and the cosmic ray ionization rate is an order of magnitude higher [14,17]. Our conclusion that the strengths of the J-band and H-band DIBs in Cyg OB2 and the GC are in rough proportion to the extinction suggests that the carriers of these bands survive equally well in both environments.

Interstellar extinction is mainly caused by silicate-based dust particles[18] and thus it is not obvious that the strengths of the 1.318 µm and 1.527 µm DIBs should depend only on the extinction in diffuse clouds if, as suspected, their carriers are carbonaceous. Indeed, in the Galactic plane the strengths of many visual wavelength DIBs do not correlate well with extinction[19] or with one another from source to source [4, 19]. Due to the high extinction towards the GC, no information on the strengths of visual DIBs in that direction is available and thus comparisons of them with the new H-band DIBs cannot be made.

In contrast to the H-band DIBs the strength of the 3.4 µm interstellar hydrocarbon absorption band is 2-3 times higher toward the GC than one would predict by scaling its strength in Galactic diffuse clouds by the ratio of extinctions[20]. The different behaviors of two families of probable carbonaceous particles, the carriers of the H-band DIBs and the carriers of the 3.4 µm feature, indicate that their abundances respond differently to the different physical conditions in and outside of the GC and suggests that the carriers are not closely coupled by interstellar chemistry. A similar conclusion has been drawn for a limited number of visual wavelength DIBs observed in several other sightlines[21]. Differences in the interstellar carbon abundance in diffuse clouds in the Galactic plane and in the GC could also affect the relative strengths of the bands in those locations, but little information on the carbon abundance in the GC is available.

## Table 1: Basic Properties of Newly Discovered DIBs[a]

| Wavelength[b] (μm) v(LSR)=0 km/s | FWHM[c,d] (Å) | $\tau_{obs}$ (max)[c] | $W_\lambda$ (Å)[e] | No. of detections[f] |
|---|---|---|---|---|
| 1.5225 | 30 ± 10 | 0.04 ± 0.01 | 1.6±0.4 | 4 |
| 1.5273 | 6 ± 1 | 0.19 ± 0.01 | 1.5±0.1 | 6 |
| 1.5617 | 10 ± 2 | 0.05 ± 0.01 |  | 5[g] |
| 1.5653 | 15 ± 4 | 0.07 ± 0.01 | 2.8±0.3[h] | 6 |
| 1.5673 | 9 ± 2 | 0.07 ± 0.01 |  | 6 |
| 1.5990 | 9 ± 2 | 0.015 ± 0.007 | 0.15±0.05 | 4 |
| 1.6232 | 24 ± 3 | 0.045 ± 0.005 | 1.3±0.3 | 6 |
| 1.6573 | - | 0.022 ± 0.006 |  | 4 |
| 1.6585 | - | 0.028 ± 0.006 | 0.56±0.12[h] | 6 |
| 1.6596 | - | 0.018 ± 0.006 |  | 5 |
| 1.7758 | 8 ± 2 | 0.05 ± 0.01 | 0.6±0.2 | 6 |
| 1.7807 | 12 ± 3 | 0.08 ± 0.01 | 1.0±0.3 | 6 |
| 1.7930 | 35 ± 15 | 0.06 ± 0.02 | 1.5±0.5 | 6 |

[a] measured toward GCS3-2
[b] uncertainty in wavelength is 0.0010 μm
[c] eyeball estimates; uncertainties depend on strength of DIB and signal-to-noise ratio in wavelength region of DIB
[d] deconvolved, assuming Gaussian DIB profile and Gaussian instrumental profile
[e] determined by numerical integration
[f] toward GC sources, out of six sightlines observed (three shown here, three unpublished)
[g] probably present in sixth star, but low S/N of spectrum and strong and broad emission lines prohibits clear detection.
[h] includes $W_\lambda$ of DIBs at adjacent wavelengths

**Acknowledgements:**
This paper is based on observations obtained at the Gemini Observatory, which is operated by the Association of Universities for Research in Astronomy, Inc., under a cooperative agreement with the NSF on behalf of the Gemini partnership: the National Science Foundation (United States), the Science and Technology Facilities Council (United Kingdom), the National Research Council (Canada), CONICYT (Chile), the Australian Research Council (Australia), Ministério da Ciência e Tecnologia (Brazil) and Ministerio de Ciencia, Tecnología e Innovación Productiva (Argentina). Financial support for this research has been given by the Spanish Ministerio de Ciencia e Innovación. We thank A. Lenorzer for her reductions of the H-band spectra of the Cygnus OB2 stars. We are grateful to B. J. McCall and T. Oka for reviewing a preliminary version of the manuscript.

**Contributions**
T.R.G. and F.N. wrote the observing proposal. T.R.G. obtained the data. B.W.S. and T.R.G. reduced the data. F.N and T.R.G. recognized the spectral features as DIBs. T.R.G., F.N., D.F.F., D.F. and B.W.S. discussed the results. T.R.G., F.N., and D.F.F. wrote the paper.

**Figure captions**

Fig. 1. Observed spectra of three hot stars in the Galactic Centre (GC) and an average spectrum of seven stars in the Cygnus OB2 association. The wavelengths of five newly discovered diffuse interstellar bands (DIBs) are indicated by vertical dotted lines. The observations of the GC sources were carried out in July 2010 on Mauna Kea at the Frederick C. Gillett Gemini North Telescope using its near-infrared integral field spectrograph NIFS. Spectra of the unreddened early A-type dwarf stars, HD 156721 and HD 174249, located outside of the GC and far from GC sightlines and observed at similar airmasses, served as telluric standards. The prominent hydrogen absorption lines in their spectra were removed by dividing by a model spectrum of the A-star Vega. The spectra of the GC objects were divided by these "de-lined" A-star spectra in order to remove telluric absorption lines. The continuum of each resultant spectrum was fitted with a spline function and divided by that function to produce the normalized spectra, shown here at R~4,000 (equivalent to a wavelength resolution of 0.00038 μm at 1.54 μm). Wavelength calibration, obtained using telluric absorption lines, is accurate to 0.0001 μm. The prominent emission lines in the spectrum of X174516 and absorption lines in the spectrum of qF362, are Brackett series transitions of atomic hydrogen (lower level quantum number 4 and upper level quantum numbers 15-20), which are often present in hot and luminous stars. The strongest DIB, at 1.5273 μm, is blended with one of the Brackett lines in the spectra of X174516 and qF362, but is uncontaminated in the spectrum of GCS3-2. The Cygnus OB2 average spectrum (R=2250) is based on individual spectra obtained in 2002 at the Telescopio Nazionale Galileo on La Palma, which were reduced using similar techniques as described above.

Fig. 2. Spectra of the newly discovered diffuse interstellar bands (DIBs). Vertical lines in each panel indicate peak wavelengths and horizontal lines the observed full widths at half maximum. Spectra are shown for three stars: GCS3-2 (continuous lines), X174516 (dashed lines), and qF362 (dotted lines). The DIBs are at the following wavelengths: 1.5225, 1.5273, 1.5617, 1.5653, 1.5673, 1.5990, 1.6232, 1.6573, 1.6585, 1.6596, 1.7758, 1.7807, and 1.7930 μm. Most are considerably weaker than the four strongest ones (shown both in Fig. 1 and in the top two panels here). In each of the remaining panels the spectrum of more than one star is shown. Since the spectra of the individual stars are diverse, we cannot rule out the presence of contaminating stellar lines in some of these spectral regions. The spectrum of GCS3-2 is the least contaminated by stellar lines; thus its spectrum has been weighted most strongly by us in deciding which absorption features are DIBs. The H-band DIBs have a wide range of spectral widths. The optical depth of the strongest DIB at 1.5273 μm is at least twice that of any of the other H-band DIBs. Three of the panels in Fig. 2 contain, centered near 1.565 μm, 1.658 μm, and 1.784 μm. The triplet near 1.658 μm appears to be partially blended and might be regarded as one DIB. However, as discussed in the text, the large velocity dispersion of the diffuse gas in which we believe the bulk of the absorptions arise may be mostly responsible for this blending.

Fig. 3. Profile of the 1.318 μm diffuse interstellar band toward qF362. The equivalent width of the feature is 3.13 Å, with an uncertainty of ±0.10 Å, compared to 0.86±0.06 Å in the Cygnus OB2 cluster star BD +40°42203, which suffers a visual extinction of 5.97 mag [20]. Scaling extinction with the equivalent width of the 1.318 μm feature results in an estimated diffuse cloud visual extinction toward qF362 of 21.7±1.0 mag, which is consistent with a previously estimate of diffuse cloud extinction toward the GC[12].

**Affiliations:**

Gemini Observatory, 670 N. A`ohoku Place, Hilo HI 96720, USA
T. R. Geballe

Centro de Astrobiología (CSIC-INTA), Ctra. Torrejón a Ajalvir km 4, 28850 Torrejón de Ardoz, Spain
F. Najarro, D. de la Fuente

Center for Detectors, Rochester Institute of Technology, Rochester, NY 14623, USA
D. F. Figer

Department of Physics and Astronomy, University of California, Los Angeles, CA 90095, USA
B. W. Schlegelmilch

**Corresponding Author**
Correspondence to: T. R. Geballe (tgeballe@gemini.edu)

**Competing financial interests**
The authors declare no competing financial interests.

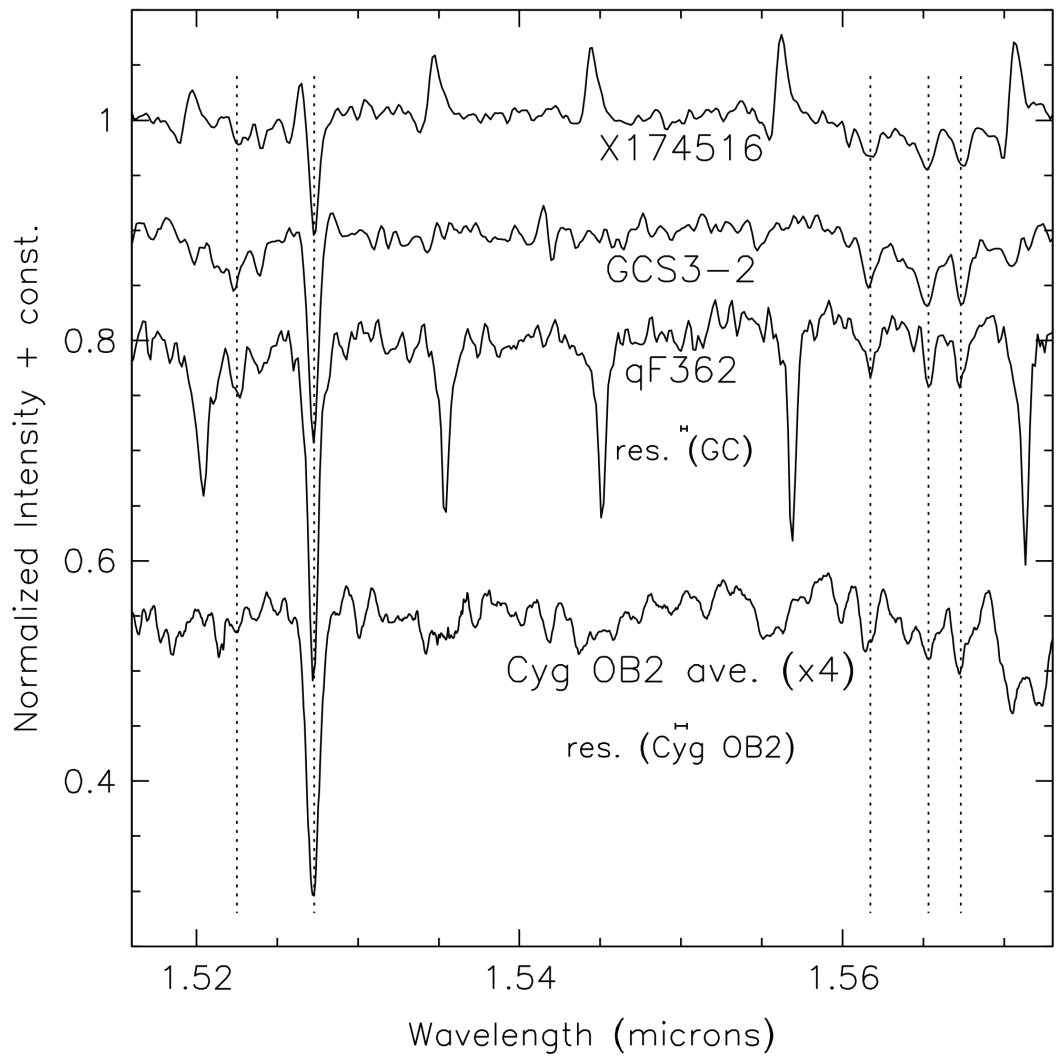

Figure 1

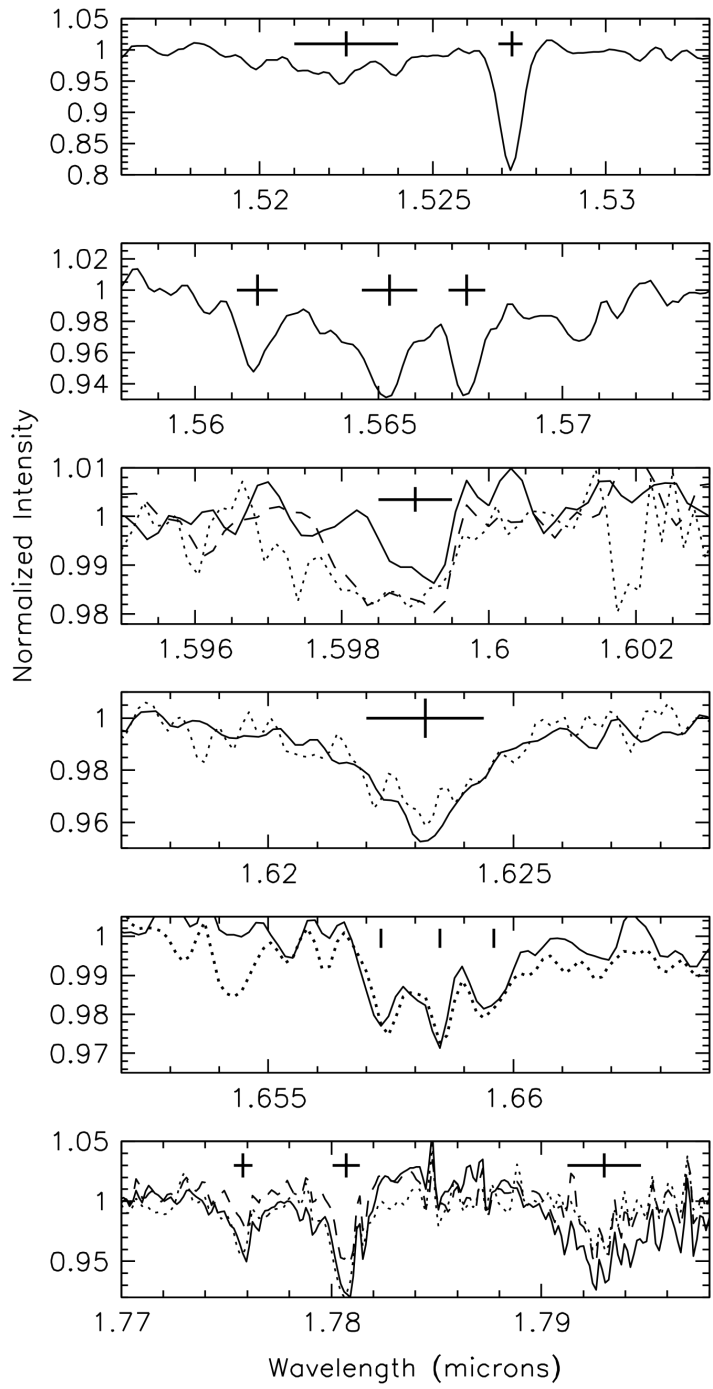

Figure 2

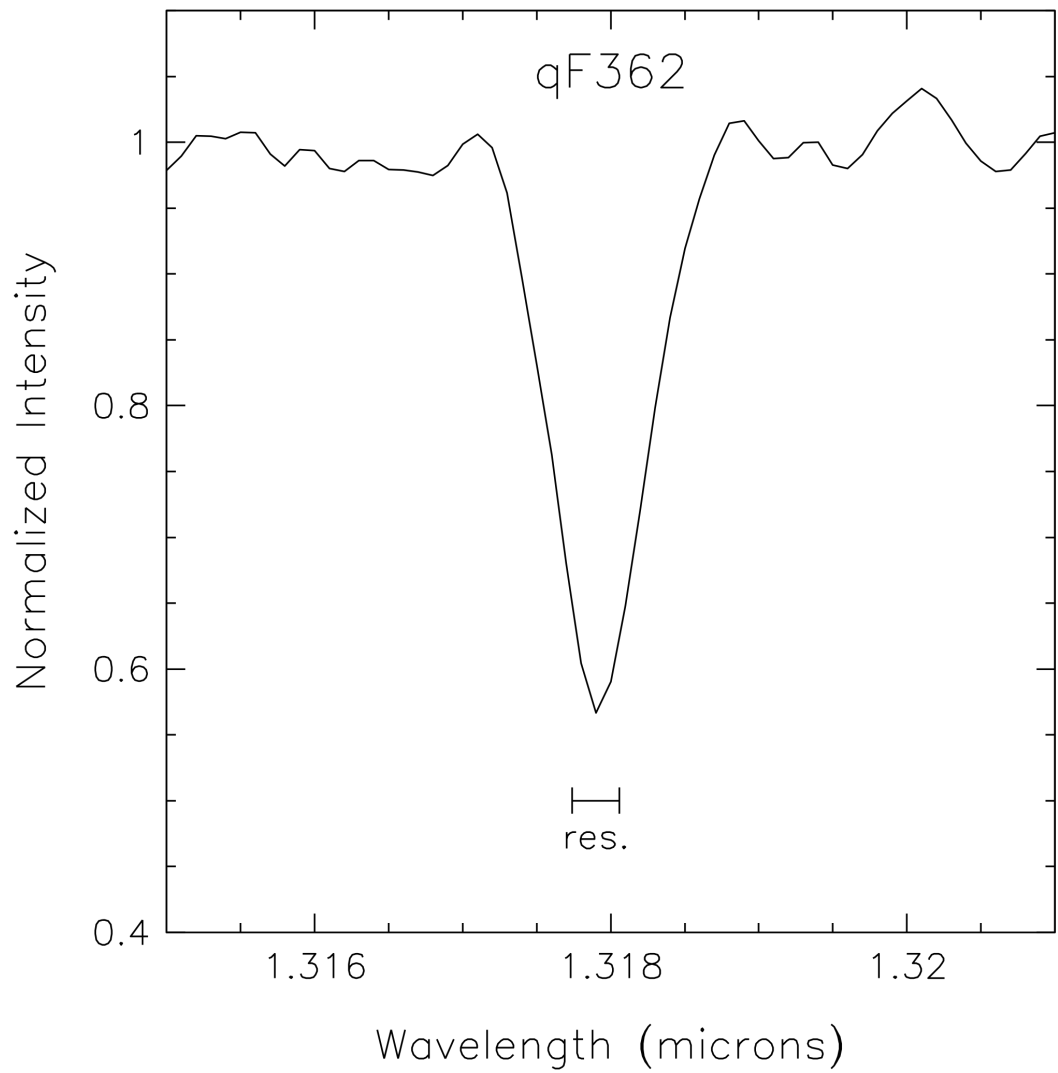

Figure 3